\documentclass{article}
 \usepackage{latexsym}
 \usepackage{amsmath}
 \usepackage{amsfonts}
\usepackage{pdflscape}  
\usepackage[pdftex]{color}
\definecolor{myred}{rgb}{0.6,0,0}
  
\newcounter{smallarabics}
\newenvironment{arabicenumerate}
{\begin{list}{{\normalfont\textrm{(\arabic{smallarabics})}}}
  {\usecounter{smallarabics}\setlength{\itemindent}{0cm}
   \setlength{\leftmargin}{5ex}\setlength{\labelwidth}{4ex}
   \setlength{\topsep}{0.75\parsep}\setlength{\partopsep}{0ex}
   \setlength{\itemsep}{0ex}}}
{\end{list}}

\newcounter{smallroman}

\newcommand{\ben}{\begin{arabicenumerate}}  
\newcommand{\een}{\end{arabicenumerate}}

\newtheorem{theoreme}{Theorem }[section]
\newtheorem{proposition}[theoreme]{Proposition}
\newtheorem{lemma}[theoreme]{Lemma}
\newtheorem{definition}[theoreme]{Definition}

\newtheorem{remark}[theoreme]{Remark}
\newtheorem{example}[theoreme]{Example}
\newtheorem{assumption}{Assumption}[section]

\def\rr{{\mathbb R}}

\def\cc{{\mathbb C}}
\def\nn{{\mathbb N}}

\def\bbbone{{\mathchoice {\rm 1\mskip-4mu l} {\rm 1\mskip-4mu l}
{\rm 1\mskip-4.5mu l} {\rm 1\mskip-5mu l}}}
\def\one{\bbbone}
\renewcommand\t{{\scriptstyle\#}}

\newcommand\Ln{{\rm Ln}}

\renewcommand\Im{{\rm Im}}
\renewcommand\Re{{\rm Re}}

\newcommand\wlim{{\rm w-}\lim}

\newcommand\Res{{\rm Res}}

\newcommand\bep{\begin{proposition}}
\newcommand\eep{\end{proposition}}
\newcommand\ber{\begin{remark}}
\newcommand\eer{\end{remark}}

\newcommand\proof{\noindent {\bf Proof.}\ \ }

\newcommand\red{}

\renewcommand\i{{\rm i}}

\newcommand\I{{\rm I}}
\newcommand\II{{\rm II}}
\newcommand\III{{\rm III}}

\newcommand{\beq}{\begin{equation}}
\newcommand{\eeq}{\end{equation}}

\newcommand{\bear}[1]{\begin{array}{#1}}
\newcommand{\ear}{\end{array}}

\renewcommand\sp{{\rm sp}}

\newcommand\qed{\hfill$\Box$\medskip}

\newcommand\cF{{\mathcal F}}

\newcommand\bel{\begin{lemma}}
\newcommand\eel{\end{lemma}}
\newcommand\bet{\begin{theoreme}}
\newcommand\eet{\end{theoreme}}
\newcommand\bex{\begin{example}}
\newcommand\eex{\end{example}}
\newcommand\bed{\begin{definition}}
\newcommand\eed{\end{definition}}
\newcommand\bea{\begin{assumption}}
\newcommand\eea{\end{assumption}}
\renewcommand\bar{\overline}

\newcommand\e{{\rm e}}

\renewcommand\d{{\rm d}}

\newcommand\Dom{\mathop{\rm Dom}}

\renewcommand\r{{\rm r}}

\newcommand\p{{\rm p}}
\usepackage{setspace}

\begin{document}

\title{Homogeneous rank one perturbations}
\author{
Jan Derezi\'{n}ski\footnote{The financial support of the National Science
Center, Poland, under the grant UMO-2014/15/B/ST1/00126, is gratefully
acknowledged. The author  thanks Serge Richard for useful discussions.}\\
Department of Mathematical Methods in Physics \\ Faculty of Physics\\
 University of Warsaw\\ Pasteura 5, 02-093, Warszawa, Poland\\
email: jan.derezinski@fuw.edu.pl}

\maketitle

\begin{abstract}
  A holomorphic family of closed operators
  with a rank one perturbation given by the function $x^{\frac{m}{2}}$ is studied.
    The operators can be used in a toy model of renormalization group.
\end{abstract}

\newpage


\section{Introduction}

Rank one perturbations can be used to illustrate various interesting mathematical concepts. For instance, they can be singular: the perturbation is not an operator and   an infinite renormalization may be needed.
Rank one perturbations are often applied to model physical phenomena.

Our paper is devoted to a special class of exactly solvable rank one perturbations, which are both singular and physically relevant.
We consider the Hilbert space  $L^2[0,\infty[$. The starting point is the operator  of multiplication by $x\in[0,\infty[$, denoted by $X$. We try to perturb it by a rank one operator
     involving the function $x^{\frac{m}{2}}$.
Thus, we try to define an operator formally given by
        \beq X+\lambda|x^{\frac{m}{2}}\rangle\langle x^{\frac{m}{2}}|.\label{oper}\eeq
        Note that we allow $m$ and $\lambda$ to be complex.
In particular,  (\ref{oper}) is usually non-Hermitian.
        The function 
          $x^{\frac{m}{2}}$ is never square integrable, and therefore, the perturbation is always singular.

(\ref{oper}) is very special. Formally, $X$ is homogeneous of degree $1$ and its perturbation is homogeneous of degree $m$. We will see that in order to define a closed operator on $L^2[0,\infty[$ one needs to restrict $m$ by the condition $-1<\Re m<1$. Besides, a special treatment is needed in the case $m=0$. One obtains two holomorphic families of closed operators, $H_{m,\lambda}$ and $H_0^\rho$. $\lambda$ and $\rho$ can be interpreted as coupling constants, which in the case $0\leq\Re m<1$ need an infinite renormalization.

    The families of the operators that we introduce are  exactly solvable in a rather strong sense: one can compute their resolvents, spectral projections and M{\o}ller (wave) operators. One can also describe their spectra, which can be quite curious.

In our opinion,  the families  $H_{m,\lambda}$ and $H_0^\rho$ that we constructed are quite instructive. 
One can argue that they provide excelent material for exercises in a semi-advanced course on operator theory, or even quantum physics. They illustrate various sophisticated concepts related to operators in Hilbert spaces (singular perturbations of various kinds, scattering theory). They can also be treated as toy models of some important ideas of theoretical physics such as renormalization group flows and breaking of scaling symmetry.

We  believe, that in some form these operators show up in many contexts in mathematics and theoretical physics, especially when we deal with scaling symmetry. Below we briefly describe one situation where
these operators are present.

As shown in \cite{BDG}, for $\Re m>-1$ one can define a holomorphic family of closed Schr\"odinger operators on  $L^2[0,\infty[$ homogeneous of degree $-2$
 given formally by
    \beq \tilde H_m=-\partial_x^2+\Big(-\frac14+m^2\Big)\frac{1}{x^2}.\label{schro}\eeq
(As compared with the notation of \cite{BDG}, we add a tilde to distinguish from the operators considered in this paper).    These operators have contiuous spectrum in $[0,\infty[$ of multiplicity $1$. They can be diagonalized
with help of the so-called Hankel transformation $\cF_m$, whose kernel has a simple expression in terms of the Bessel function $J_m$.

    As shown in \cite{DR}, for $-1<\Re m<1$ there exists a two-parameter holomorphic family of closed  operators that can be associated with the differential expression on the right hand side of (\ref{schro}). They correspond to mixed boundary conditions at zero and are denoted $\tilde H_{m,\kappa}$. The case $m=0$ needs special treatment, and one introduces a family of $\tilde H_0^\nu$. As we show in our
    paper, the operators 
    $\tilde H_{m,\kappa}$, resp. $\tilde H_0^\nu$, are equivalent (similar) to the operators
    $H_{m,\lambda}$ and $H_0^\rho$, {\red where $\kappa$ and $\nu$ are linked by a simple relation with $\lambda$ and $\rho$, see Thms \ref{link1} and \ref{link2}.}

    The operators $\tilde H_{m,\kappa}$ and $\tilde H_0^\nu$ 
    are very well motivated---they constitute natural classes of Schr\"odinger operators, which are relevant for many problems in mathematical physics. However, their theory looks complicated--it requires the knowledge of some special functions, more precisly, Bessel-type  functions and the Gamma function.
    On the other hand, the theory 
    of     $H_{m,\lambda}$ and $H_0^\rho$ does not involve special functions at all---it uses   only
trigonometric functions and the logarithm.

The paper is organized as follows. In Section
\ref{General theory of rank one perturbations}
we recall the theory of sigular rank one perturbations. It is sometimes called the Aronszajn-Donoghue theory and goes back to \cite{A,D,BF}. It is described in particular in \cite{AK,DF,KS}. We discuss also the scattering theory in the context of rank one perturbations. {\red Here the basic reference is \cite{Y}.

  Note, however, that  we do not assume that the perturbation is self-adjoint, and most of the
  literature on this subject is  restricted to the self-adjoint case.
A notable exception are the articles \cite{K1,K2}, where non-self-adjoint perturbations of self-adjoint operators are studied.}

Section 
\ref{A family of toy models} is the main part of our paper. Here we construct and study the operators
$H_{m,\lambda}$ and $H_0^\rho$.

In Section
\ref{Equivalence}
we describe the relationship of the operators
$H_{m,\lambda}$ and $H_0^\rho$ with Schr\"odinger operators with inverse square potentials  $\tilde H_{m,\kappa}$ and $\tilde H_0^\nu$. There exists large literature for such Schr\"odinger operators, see eg. \cite{GTV}, it is however usually restricted to the self-adjoint case.
The  general case is studied in \cite{BDG} and especially \cite{DR}.

In the appendix we collect some integrals that are used in our paper.


\section{General theory of rank one perturbations}
\label{General theory of rank one perturbations}

\label{Theory of rank one perturbations}
\setcounter{equation}{0}

\subsection{Preliminaries}

We consider the Hilbert space $ L^2[0,\infty[$
    with the scalar product
\begin{equation}
( f|g):=\int_0^\infty \bar{f(x)}g(x)\d x.\label{sesqui}
\end{equation}
In addition, it is also equipped with the bilinear form
\begin{equation}\label{bili}
\langle f|g\rangle:=\int_0^\infty f(x)g(x)\d x,
\end{equation}
{\red Thus we use round brackets for the sesquilinear scalar product and angular brackets for the closely related bilinear form. Note that in some sense the latter plays a more important role in our paper (and in similar exactly solvable problems) than the  former.}

If $B$ is an operator  then $B^*$ denotes the usual
Hermitian adjoint of $B$, whereas
$B^\t$  denotes {\red the {\em transpose} of $B$, that is, its adjoint w.r.t.~the
\eqref{bili}.} Clearly, if $B$ is a bounded linear operator  with
\begin{equation*}
\big(B f\big)(k):=\int_0^\infty B(k,x)f(x)\d x,
\end{equation*}
then
\begin{equation*}
\big(B^* f\big)(x)=\int_0^\infty\bar{ B (k,x)}f(k)\d k,
\end{equation*}
while
\begin{equation*}
\big(B^\t f\big)(x)=\int_0^\infty B (k,x)f(k)\d k.
\end{equation*}

An operator $B$ is self-adjoint if $B=B^*$. We will say that it is {\em
  \red self-transposed} if
\beq B^\#=B.\label{symme}\eeq
{\red It is useful to note that a holomorphic function of a self-transposed operator is self-transposed.}

{\red It is convenient to use sometimes Dirac's ``bra-ket'' notation.  For a function $f\in L^2[0,\infty[$, we have the operator
       $|f\rangle:\cc\to L^2[0,\infty[$
          given by
          \beq \cc\ni z\mapsto |f\rangle z:=zf\in  L^2[0,\infty[\label{dirac1}\eeq
              and its transpose
              $\langle f|:=|f\rangle^\#
              :L^2[0,\infty[\to \cc$
                  given by
                  \beq  L^2[0,\infty[\ni v\mapsto\langle f|v=\int f(x) v(x)\d x.\label{dirac2}\eeq

                      We will also use the same notation in the case $f$ is not square integrable---then $\langle f|$ is an unbounded operator and
                       $|f\rangle$ is an unbounded form with appropriate domains.

                      Note that (\ref{dirac1}) and (\ref{dirac2}) are consistent with the notation for 
(\ref{bili})---both use angular brackets.                                            In principle, we could also use the Dirac's bras and  kets suggested by the scalar product (\ref{sesqui}), involving round brackets,
                      \beq |f):=|f\rangle,\quad (f|:=\langle\bar f|,\eeq
                      but we prefer to use the notation associated with
      (\ref{bili}).   
          }

\subsection{Construction}
    
Let $X$ denote  the (unbounded) operator on $ L^2[0,\infty[$
given by
\begin{align}
  Xv(x)&:=xv(x),\\
  \quad v&\in\Dom(X)=\Big\{v\in L^2[0,\infty[\quad\big|\quad \int|v(x)|^2x^2\d x<\infty\Big\}.
\label{rew}\end{align}

Let $h_2,h_1$ be measurable functions on $[0,\infty[$.
Consistently with the notation introduced in (\ref{dirac1}) and (\ref{dirac2}), we will write $|h_2\rangle\langle h_1|$ for the (possibly unbounded) quadratic form given by
\beq\big(w|h_2\rangle\langle h_1|v\big):=\int \bar{w(x)}h_2(x)\d x\int h_1(y)v(y)\d
y,\eeq
for $w,v$ in the (obvious) domain of $|h_2\rangle\langle h_1|$.
(Note the absence of the complex conjugation on $h_1$).

It is well known that in some situations
\beq
H_\lambda:=X+\lambda|h_2\rangle\langle h_1|\label{aronszajn}\eeq
can be interpreted as an operator, possibly after an appropriate renormalization of the coupling constant $\lambda$. This is sometimes called the Aronszajn-Donoghue theory, and is described e.g. in  \cite{AK,DF,KS}.
We will need a somewhat non-standard version of this theory, because our rank one perturbation
does not have to be Hermitian. Therefore, we describe it in some detail.

One can consider three cases of the Aronszajn-Donoghue theory,
 with an increasing level of difficulty.
The first case is elementary:\\

\medskip

{\bf Assumption\ \ I.}\ \ $ h_1,h_2\in L^2.$\\
\medskip

Then $|h_2\rangle\langle h_1|$ is a  bounded operator. Therefore,
$H_\lambda$ is well defined on $\Dom(X)$,
and we can easily compute the resolvent of $H_\lambda$.
In fact, define
\begin{eqnarray}
  \label{adgeq1}
g_\infty (z)&:=&\langle h_1|(z-X)^{-1}|h_2\rangle=\int_0^\infty
h_1(x)(z-x)^{-1}h_2(x)\d x,\\
\label{adgeq2}
g_\lambda (z)&=&-\lambda^{-1}+g_\infty(z).
\end{eqnarray}
Then
\beq
\sp H_\lambda\subset\{z\ :\ g_\lambda(z)=0\}\cup[0,\infty[,\eeq
    and for such $z$ such that $g_\lambda(z)\neq0$, $z\not\in[0,\infty[$,
\beq(z-H_\lambda)^{-1}=R_\lambda(z),\label{resi}
\eeq
where
\begin{alignat}{4}
R_\lambda (z)
&=&&(z-X)^{-1}\\&&&-g_\lambda(z)^{-1}(z-X)^{-1}|h_2\rangle\langle
h_1|(z-X)^{-1},\quad\lambda\neq0;
\label{krei0}\\
\label{adrezeq1} R_0(z)&:=&&(z-X)^{-1}.\end{alignat}

Consider now \\

\medskip

{\bf
  Assumption\ \ II.} \ \  $\frac{h_1}{1+X},\frac{h_2}{1+X}\in
L^2,\ \ \ \frac{h_1h_2}{1+X}\in L^1$.\\

\medskip

  Then it is easy to check that $g_\lambda$ and
  $R_\lambda(z)$ are  still well defined.
  Besides, $R_\lambda(z)$ satisfies the resolvent equation,
has zero kernel and dense range. Hence, by the theory of
pseudoresolvents \cite{Kato2},
there exists a closed operator
$H_\lambda$ such that
(\ref{resi}) is true. Note that $H_0=X$
  and often we can include $\lambda=\infty$.

    Finally, consider\\

    \medskip

{\bf
  Assumption\ \ III.} \ \ $\frac{h_1}{1+X},\frac{h_2}{1+X}\in
L^2$.\\
\medskip

Then $g_\lambda$ is in general ill defined. Instead, we consider the equation
\beq\partial_zg(z)=-\langle h_1|(z-X)^{-2}|h_2\rangle.\label{adgeq3}\eeq
If one of solutions of (\ref{adgeq3}) is called $g^0$, then all other are
given by
\beq g^\rho(z):=\rho+ g^0(z),
\eeq
for some $\rho\in\cc$.
We set
\begin{alignat}{4}
R^\rho (z)
&:=&&(z-X)^{-1}\\&&&-g^\rho(z)^{-1}(z-X)^{-1}|h_2\rangle\langle h_1|(z-X)^{-1},
\label{krei0a}\\
\label{adrezeq1a} R^\infty(z)&:=&&(z-X)^{-1}.\end{alignat}
Again, $R^\rho(z)$ is a pseudoresolvent and by \cite{Kato2}
there exists a unique family of operators $H^\rho$ such that
\beq (z-H^\rho)^{-1}=R^\rho(z).\eeq

We have thus constructed a family of operators. Under Assumption I or
II, it can be written as $H_\lambda$, where
$\lambda\in\cc\cup\{\infty\}$ has the meaning of a coupling
constant. Under Assumption III, in general, the coupling constant may
lose its meaning, and we may be forced to use the parametrization
$H^\rho$, where again $\rho\in\cc\cup\{\infty\}$. (In practice,
however, as we will see,
the notation $H_\lambda$ could be natural even if Assumption
II does not hold).

\subsection{Point spectrum}

Untill the end of this section we suppose that
Assumption III is satisfied. We consider an operator $H$ of the form  $H_\lambda$ or $H^\rho$, as described above. Thus,
\begin{alignat}{4}
(z-H)^{-1}
&=&&(z-X)^{-1}\\&&&-g(z)^{-1}(z-X)^{-1}|h_2\rangle\langle h_1|(z-X)^{-1},
\label{krei0b}\end{alignat}
where $g=g_\lambda$ or $g=g^\rho$.

It is easy to see that the spectrum of $H$ consists of $[0,\infty[$ and  eigenvalues
at
\beq\{w\in\cc\backslash[0,\infty[\ :\ 0=g(w)\}.\eeq
If {\red $w\in\cc\backslash[0,\infty[$ is an eigenvalue with $\langle h_1|(w-X)^{-2}|h_2\rangle\neq0$, then it is simple} and the corresponding eigenprojection is given by the formula
    \beq\one_{\{w\}}(H)
    =\frac
    {(w-X)^{-1}|h_2\rangle\langle h_1|(w-X)^{-1}}
    {\langle h_1|(w-X)^{-2}|h_2\rangle}.\eeq

  \subsection{Dilations}\label{Dilations}
Before we continue, let us say a few words about
the group of dilations
\beq U_\tau f(x);=\e^{\frac\tau2}f(\e^\tau x),\ \ \tau\in\rr.\eeq
It can be written as $U_\tau=\e^{\i\tau A}$, where
the generator of dilations is
 \beq A:=\frac1{2
   \i}(x\partial_x+\partial_x x).\eeq
    We say that $H$ is homogeneous of degree $p$ if
    $U_\tau H U_\tau^*=\e^{pt}H$. For instance, $X$ is homogeneous of
    degree $1$.

It is easy to see that  an operator $B$ on $L^2(\rr_+)$ has the integral kernel
    \beq B(x,y)=\frac{1}{\sqrt{xy}}\phi\Big(\ln\frac{x}{y}\Big),\eeq
    iff $B=\hat\phi(A)$, where
    \beq\hat\phi(\xi)=\int\phi(t)\e^{-\i t\xi}\d t,\label{melin}
    \eeq
    see e.g. \cite{BDG}. For example,  by (\ref{integral1}), \beq
    \frac{\pm2\pi\i}{(\e^{\pm 2\pi A}+\one)}\quad\text{  has the  kernel }\quad\frac{
  1}{(x-y\mp\i0)}.\label{pap}\eeq

    \subsection{Essential spectrum}

It follows from the Weyl Theorem that the essential spectrum of
$H$ is $[0,\infty[$. 
Detailed study of the essential spectrum requires technical
assumptions on the perturbation \cite{Y}. In this section, we will limit
ourselves to a heuristic theory, without specifying precise  assumptions.
It will be possible to justify these formulas in
 the concrete situation considered in our paper.

First, we will check that  the kernel of $    \one_{[0,\infty[}(H)$, that is, of the spectral projection of $H$ onto
$[0,\infty[$ is given by the following formula:
    \begin{eqnarray}\label{kerno}
    \one_{[0,\infty[}(H)(x,y)
        &=&
        \delta(x-y)\\
        &&\hspace{-6ex}+\frac{h_2(x)h_1(y)}{(x-y-\i0)}\Big(
        \frac{1}{g(y+\i0)}-\frac{1}{g(x+\i0)}\Big)\\
        &&
\hspace{-22ex}        +\frac{1}{2\pi\i}\frac{h_2(x)h_1(y)}{(x-y-\i0)}\int\d s
        \frac{h_1(s)h_2(s)}{g(s+\i0)g(s-\i0)}
        \Big(\frac{1}{(x-s-\i0)}+\frac{1}{(s-y-\i0)}\Big).\nonumber
    \end{eqnarray}

    To see this we use the Stone formula
    \beq
    \one_{[0,\infty[}(H)=\wlim_{\epsilon\searrow0}
        \frac{1}{2\pi\i}\int_0^\infty\d s
        \Big(        (s-\i\epsilon-H)^{-1}- (s+\i\epsilon-H)^{-1}\Big).
        \label{kerne}\eeq
       Thus
        \begin{eqnarray*}
          &&\one_{[0,\infty[}(H)(x,y)\\
              &=&\frac{1}{2\pi\i}\int_0^\infty\d s \Big(\frac{1}{(s-\i0-x)}\delta(x-y)
-     \frac{1}{(s+\i0-x)}\delta(x-y)
             \Big)\\ 
          &&\hspace{-5ex}+\frac{1}{2\pi\i}\int_0^\infty\hspace{-2ex}\d
          s\Big(
\frac{ h_2(x)h_1(y)}{g(s-\i0)(s-\i0-x)(s-\i0-y)}
-\frac{h_2(x)h_1(y)}{g(s+\i0)(s+\i0-x)(s+\i0-y)}\Big)
\nonumber\\
&=&\int_0^\infty\d s\delta(s-x)\delta(x-y)\\
&&+
\frac{1}{2\pi\i}\int_0^\infty\d      s\frac{h_2(x)h_1(y)}{g(s-\i0)(s-\i0-x)}\Big(\frac{1}{(s-\i0-y)}-\frac{1}{(s+\i0-y)}\Big) 
          \nonumber
          \\
&&+ \frac{1}{2\pi\i}\int_0^\infty\d
          s\Big(\frac{1}{(s-\i0-x)}-
      \frac{1}{(s+\i0-x)}\Big)\frac{h_2(x)h_1(y)}{(s+\i0-y)g(s+\i0)}\nonumber\\
&&+  \frac{1}{2\pi\i}\int_0^\infty\d
          s\Big(\frac{1}{g(s-\i0)}-\frac{1}{g(s+\i0)}\Big)
 \frac{h_2(x)h_1(y)}{(s-\i0-x)(s+\i0-y)}\nonumber\\
  \nonumber\\
  &=&\delta(x-y)\\
&&  -\int_0^\infty\d
          s\Big(\frac{h_2(x)h_1(y)}{g(s-\i0)(s-\i0-x)}\delta(s-y) 
          +\delta(s-x)\frac{h_2(x)h_1(y)}{(s+\i0-y)g(s+\i0)}\Big)\nonumber\\
&&+  \frac{1}{2\pi\i}\int_0^\infty\d
          s\frac{h_1(s)h_2(s)}{g(s-\i0)g(s+\i0)}
          \frac{h_2(x)h_1(y)}{(x-y+\i0)}\Big(\frac{1}{(s-\i0-x)}-
          \frac{1}{(s+\i0-y)}\Big).\nonumber
        \end{eqnarray*}

\subsection{M{\o}ller operators I}
\label{Moller operators}

{\red For the purpose of our paper, we define the M{\o}ller operators between $X$ and $H$, denoted $W^\pm(H,X)$ and $ W^\pm(X,H)$, by describing their kernels:
\begin{eqnarray}
  W^\pm(H,X;x,y)&=&\delta(x-y)+
  \frac{h_2(x)h_1(y)}{(x-y\pm\i0)g(y\mp\i0)},\label{easi}\\
    W^\pm(X,H;x,y)&=&\delta(x-y)+
    \frac{h_2(x)h_1(y)}{g(x\pm\i0)(y-x\mp\i0)}.\label{easi1}\end{eqnarray}
 To motivate the definitions (\ref{easi}) and (\ref{easi1})
recall that in the literature on stationary scattering theory, e.g. \cite{Y}, the M{\o}ller operators are often  introduced as follows:
\begin{eqnarray}
  W^\pm(H,X)&:=&\wlim_{\epsilon\searrow0}\frac{\epsilon}{\pi}\int\d
  s(s\mp\i\epsilon-H)^{-1}
  (s\pm\i\epsilon-X)^{-1},\label{stat1}\\
     W^\pm(X,H)&:=&\wlim_{\epsilon\searrow0}\frac{\epsilon}{\pi}\int\d
  s(s\mp\i\epsilon-X)^{-1}
  (s\pm\i\epsilon-H)^{-1}
.\label{stat2}  \end{eqnarray}
 If (\ref{stat1}) and (\ref{stat2}) exist, then a formal computation shows that their kernels are given by (\ref{easi}) and (\ref{easi1}).

In general, there is no guarantee that $W^\pm(H,X)$ and $ W^\pm(X,H)$ exist as bounded operators. If this is the case, we expect the following properties:
\begin{eqnarray}
  W^\pm(X,H)  W^\pm(H,X)&=&\one,\label{p1}\\
    W^\pm(H,X) W^\pm(X,H)&=&\one_{[0,\infty[}(H),\label{p2}\\
    W^\pm(H,X) X&=&H W^\pm(H,X).\label{p3}
\end{eqnarray}

 A rigorous derivation of (\ref{p1}), (\ref{p2}) and (\ref{p3}) for some classes of perturbations can be found in \cite{Y}. It is not very difficult to derive these identities on a formal level.

Let us give a formal derivation of (\ref{p1}):
\begin{align}
  &W^\pm(X,H)W^\pm(H,X)(x,y)\label{drop}\\
   =&\delta(x-y)
    +   \frac{h_2(x)h_1(y)}{g_\lambda(x\pm\i0)(y-x\mp\i0)}
   +  \frac{h_2(x)h_1(y)}{(x-y\pm\i0)g_\lambda(y\mp\i0)}
\\&+\int\d t
    \frac{h_2(x)h_1(t)    h_2(t)h_1(y)}{g_\lambda(x\pm\i0)(t-x\mp\i0)(t-y\pm\i0)g_\lambda(y\mp\i0)}.
\end{align}
Now
\begin{align}
  &\int\d t
  \frac{h_1(t)    h_2(t)}{(t-x\mp\i0)(t-y\pm\i0)}
  \\
  =&
  \int\d t  \frac{h_1(t)    h_2(t)}{(y-x\mp\i0)}\Bigg(-\frac{1}{(t-x\mp\i0)}+\frac{1}{(t-y\pm\i0)}\Bigg)
    \\
    =& \frac{1}{(y-x\mp\i0)}\big(-g_\lambda(x\pm\i0)+g_\lambda(y\mp\i0)\big).
\end{align}
Therefore, (\ref{drop}) is $\delta(x-y)$.

We omit the derivation of (\ref{p2}), which  is similar to that of (\ref{p1}), although somewhat more difficult, since we need to use (\ref{kerno}).

To obtain (\ref{p3}), we will compute that
\begin{eqnarray}
     W^\pm(H,X) (z-X)^{-1}&=&(z-H)^{-1} W^\pm(H,X).\label{p3-}
    \end{eqnarray}
Indeed,
\begin{align}
  &(z-H)^{-1} W^\pm(H,X)(x,y)\label{e1}\\
  =&\frac{\delta(x-y)}{(z-x)}-\frac{h_2(x)h_1(y)}{g_\lambda(z)(z-x)(z-y)}\label{e2}\\
  &+\frac{h_2(x)h_1(y)}{(z-x)(x-y\pm\i0)g_\lambda(y\mp\i0)}\label{e3}\\
  &-\int\frac{h_2(x)h_1(t)h_2(t)h_1(y)}
  {g_\lambda(z)(z-x)(z-t)(t-y\pm\i0)g_\lambda(y\mp\i0)}\d t\label{e4}\\
    =&\frac{\delta(x-y)}{(z-y)}-\frac{h_2(x)h_1(y)}{g_\lambda(z)(z-x)(z-y)}\label{e5}\\
    &+\frac{h_2(x)h_1(y)}{(z-x)(z-y)g_\lambda(y\pm\i0)}\label{e6}\\
    &+\frac{h_2(x)h_1(y)}{(x-y\pm\i0)(z-y)g_\lambda(y\mp\i0)}\label{e7}\\
  &-\int\frac{h_2(x)h_1(t)h_2(t)h_1(y)}
    {g_\lambda(z)(z-x)(z-t)(z-y)g_\lambda(y\mp\i0)}\d t\label{e8}\\
      &-\int\frac{h_2(x)h_1(t)h_2(t)h_1(y)}
  {g_\lambda(z)(z-x)(t-y\pm\i0)(z-y)g_\lambda(y\mp\i0)}\d t.\label{e9}
\end{align}
To handle (\ref{e8})${+}$(\ref{e9}) we note that
\beq
\int\frac{h_1(t)h_2(t)}{(z-t)}\d t-
\int\frac{h_1(t)h_2(t)}{(y-t\mp\i0)}\d t=g_\lambda(z)-g(y\mp\i0).\eeq
Therefore, (\ref{e8})${+}$(\ref{e9}) cancels the second term of (\ref{e5}) and (\ref{e6}). Thus, (\ref{e1}) equals
\begin{align}
  &\frac{\delta(x-y)}{(z-y)}
    +\frac{h_2(x)h_1(y)}{(x-y\pm\i0)(z-y)g_\lambda(y\mp\i0)}\label{e10}\\
  =&W^\pm(H,X)(z-X)^{-1}(x,y),\end{align}
which proves (\ref{p3-}).}

\subsection{M{\o}ller operators II}

Let us now consider $H=H_\lambda$.

We can rewrite (\ref{easi}) in terms of $W^\pm(H_0,X)=\one$ and
$W^\pm(H_\infty,X)$:
\begin{eqnarray}\notag
W^\pm(H_\lambda,X)&
=&\frac{1}{\big(1-\lambda g_\infty(X\mp\i0)\big)}\\
&&-W^\pm(H_\infty,X)
\frac{\lambda g_\infty(X\mp\i0)}{\big(1-\lambda g_\infty(X\mp\i0)\big)}
\label{formu}\end{eqnarray}

Then we consider $H^\rho$.
In general, there is no analog of (\ref{formu}). Instead, using (\ref{pap}) one obtains the following
 compact formula for $ W^{\pm}(H^\rho,X)$:
\beq  W^{\pm}(H^\rho,X)
=\one\mp h_2(X)\frac{2\pi\i}{(\e^{\pm2\pi A}+\one)}\frac{h_1(X)}{
\big(\rho+g^0(X\mp\i0)\big)}.
\label{compact}\eeq

\subsection{{\red Self-transposed}  and self-adjoint cases}

In applications, it often happens that one of the following two
conditions holds:
\begin{eqnarray} h_1=h_2=:h,&
\hbox{ resp. }& \bar h_1=h_2=:h,\label{sym2}\end{eqnarray}
(The former is the case of our paper; the latter is in most of the literature). Then
Assumptions I, II, III slightly simplify {\red and can be rewritten in terms of the scale of Hilbert spaces associated with the positive operator $1+X$:\\
{\bf Assumption I.}  $ h\in L^2[0,\infty[.$\\
{\bf Assumption II.}  $ h\in (1+X)^{\frac12}L^2[0,\infty[.$\\
{\bf Assumption III.}  $h\in (1+X) L^2[0,\infty[.$\\
} Moreover,
the family
$H_\lambda$ is {\red self-transposed}, resp. self-adjoint, that is
\[H_\lambda^\t=H_\lambda,\ \hbox{ resp. }\ H_\lambda^*=H_\lambda,\]
and the M{\o}ller operators satisfy
{\red\beq\big(W^\pm(H,X)\big)^\t=W^\mp(X,H),\ \hbox{ resp.} \
\big(W^\pm(H,X)\big)^*=W^\pm(X,H)
.\label{moller}\eeq
To see (\ref{moller}) it is enough to look at the kernels
(\ref{easi}) and (\ref{easi1}).}

\section{Family of rank one perturbations}
\label{A family of toy models}
\subsection{Construction}
\setcounter{equation}{0}

We still consider $L^2[0,\infty[$ with the operator $X$ defined as
    in (\ref{rew}).
    Let $m\in\cc$ and
    \[h_m(x):=x^{\frac{m}{2}}.\]
    We would like to define an operator formally given by
    \beq
    H_{m,\lambda}:=X+\lambda|h_m\rangle\langle h_m|.\label{toy}\eeq
    We check that for $-1<\Re m<0$ Assumption II is satisfied.
    Therefore  the construction described in Section
    \ref{Theory of rank one perturbations} allows us
    to define a closed operator $H_{m,\lambda}$ for $m$ in this range
    and $\lambda\in\cc\cup\{\infty\}$. {\red Using (\ref{app4}),} we
    compute for $z\in\cc\backslash[0,\infty[$:
    \begin{eqnarray}
&&\langle
h_m|(z-X)^{-1}|h_m\rangle^{-1}\\&=&\int_0^\infty x^{m}(z-x)^{-1}\d
x\\
&=&-(-z)^{m}\int_0^\infty\Big(\frac{x}{-z}\Big)^{m}\Big(1+\frac{x}{(-z)}\Big)^{-1}
\frac{\d
  x}{(-z)}\\
&=&(-z)^{m}\frac{\pi}{\sin \pi m}.
    \end{eqnarray}

    Next note that Assumption III is satisfied for $-1<\Re
    m<1$. Moreover, equation
    \[\partial_z g(z)= -\langle
    h_m|(z-X)^{-2}|h_m\rangle^{-1}=-(-z)^{m-1}\frac{\pi m}{\sin\pi m}\]
    can be solved, obtaining the following solutions
    \begin{eqnarray}
    g_\lambda(z)&:=&-\lambda^{-1}+(-z)^{m}\frac{\pi}{\sin \pi
      m},\ \ m\neq0,\\
    g^\rho(z)&=&\rho-\ln(-z),\ \ \ \ \ \ \ \ \ \ \ \ \ \ \ \ \ \ m=0.\end{eqnarray}
    Note that only for $m=0$ we use the ``superindex notation'' and we do not use the ``coupling constant'' $\lambda$. For $m\neq0$ we
    keep the ``coupling constant'' $\lambda$, even though for $\Re
    m\geq0$ it has lost its meaning described by (\ref{toy}).
    
    The following theorem summarizes our construction:

    \bet\ben\item For any $-1<\Re m<1$,
    $m\neq0$, $\lambda\in\cc\cup\{\infty\}$,
    there exists a unique closed operator $H_{m,\lambda}$ such that
\begin{eqnarray}\label{reso1}
  (z-H_{m,\lambda})^{-1}&=&(z-X)^{-1}\\&&\hspace{-6ex}+\Big(\lambda^{-1}-(-z)^{m}\frac{\pi}{\sin\pi
    m}\Big)^{-1} (z-X)^{-1}|h_m\rangle\langle
  h_m|(z-X)^{-1}.\nonumber
\end{eqnarray}   In particular, $H_{m,0}=X$.
\item For any $  \rho\in\cc\cup\{\infty\}$,
    there exists a unique closed operator $H_0^\rho$ such that
  \begin{eqnarray}
    (z-H_{0}^\rho)^{-1}&=&(z-X)^{-1}\\&&-\big(\rho+\ln (-z)\big)^{-1} (z-X)^{-1}|h_0\rangle\langle
  h_0|(z-X)^{-1}.\nonumber
  \end{eqnarray}
  In particular, $H_0^\infty=X$.
\een\eet

Note that the operators $H_{m,\lambda}$ and $H_0^\rho$ are {\red self-transposed.} 

It will be convenient to introduce
 the shorthand
\begin{equation}\varsigma(m,\lambda)=\varsigma:=
\lambda\frac{\pi}{\sin\pi m}.
\label{var}\end{equation}
Then we can rewrite (\ref{reso1}) as
\begin{eqnarray}
  (z-H_{m,\lambda})^{-1}&=&\frac{1}{1-\varsigma(-z)^m}(z-X)^{-1}\\&&
  -\frac{\varsigma(-z)^m}{1-\varsigma(-z)^m}(z-H_{m,\infty})^{-1}.
\nonumber\end{eqnarray}

It is possible to include both $H_{m,\lambda}$ and $H_0^\rho$ in a
single analytic family of closed operators  (see \cite{Kato2}).
\bet For $m\neq0$, set $\lambda(m,\rho):=\frac{m}{1-m\rho}$. Then
\beq (\rho,m)\mapsto\begin{cases}H_{m,\lambda(m,\rho)},& m\neq0;\\
H_0^\rho,&m=0;\end{cases}\label{annal}
\eeq
is an analytic family (their resolvents depend analytically on $(\rho,m)$).
\label{annal1}
\eet

\proof
We have
\beq\rho(m,\lambda):=m^{-1}-\lambda^{-1}.\label{invi}
\eeq
Hence, for small $m$
\begin{eqnarray}
g_\lambda(z)&=
&-\lambda^{-1}+\e^{m\ln(-z)}\frac{\pi}{\sin \pi
      m}\\
&=&-\lambda^{-1}+\big(1+m\ln(-z)\big)m^{-1}+O(m)\\
&=&\rho(m,\lambda)+\ln(-z)+O(m).
\end{eqnarray}
\qed

\subsection{Toy model of the renormalization group}

The group of dilations (``the renormalization group'') acts on our
operators in a simple way:
\begin{eqnarray}
  U_\tau H_{m,\lambda}U_\tau^{-1}&=&\e^\tau H_{m,\e^{\tau
      m}\lambda},\\
  U_\tau H_{0}^\rho U_\tau^{-1}&=&\e^\tau H_0^{\rho+\tau}.
\end{eqnarray}

We will show that an appropriately renormalized operator of the form $X$ plus a rather arbitrary  rank one perturbation is driven by the scaling to one of the operators that we consider in our paper.

\bet
Suppose that $h\in L^2[0,\infty[$ has a compact support and $h=x^{\frac{m}{2}}$ close to
$x=0$. Set
\beq H(\lambda):=X+\lambda
|h\rangle\langle h|.\eeq
We then have the following statements,
(where $\lim$ denotes the norm resolvent limit):
\ben
\item For $-1<\Re m<0$, 
\beq  \lim_{\tau\to-\infty}\e^{-\tau}U_\tau H(\lambda\e^{-m\tau})
U_\tau^{-1}=H_{m,\lambda}.\eeq
\item
For $m=0$,
\beq  \lim_{\tau\to-\infty}\e^{-\tau}U_\tau H\Big(\frac{1}{\tau}\Big)
U_\tau^{-1}=H_{0}^\nu,\eeq
where
\begin{eqnarray}
  \nu:&=&\int_0^\infty\ln(y)(h^2)'(y)\d y.\end{eqnarray}
\item  For $0\leq \Re m<1$, $m\neq0$, 
\beq  \lim_{\tau\to-\infty}\e^{-\tau}U_\tau
H\Big(\frac{\lambda}{\e^{m\tau }-\alpha\lambda}\Big)
U_\tau^{-1}=H_{m,\lambda},\eeq
where
\beq
\alpha:=\int_0^\infty\frac{h^2(x)}{x}\d x.\eeq
  \een
  \eet

\proof
We will prove only the case $0\leq \Re m<1$, $m\neq0$. The other cases are easier.
Set $\lambda_\tau:=
\frac{\lambda}{\e^{m\tau }-\alpha\lambda}$.
\begin{eqnarray*}
  &&\big(z-\e^{-\tau}U_\tau H(\lambda_\tau)U_{-\tau}\big)^{-1}\\
  &=&\e^\tau U_\tau\big(z\e^\tau-H(\lambda_\tau)\big)^{-1}U_{-\tau}\\
  &=&\e^\tau U_\tau\big(z\e^\tau-X\big)^{-1}U_{-\tau}\\
  &&-\Big(-\lambda_\tau^{-1}+\int_0^\infty h(x)^2(z\e^\tau-x)^{-1}\d
  x\Big)^{-1}\\
  &&\times \e^{\tau}U_\tau(z\e^\tau-X\big)^{-1}|h\rangle\langle h|
(z\e^\tau-X\big)^{-1}
  U_{-\tau}\\
  &=&\I-\II^{-1}\times \III.
\end{eqnarray*}
Clearly,
\begin{eqnarray*}
  \I&=&(z-X)^{-1},\\
  \e^{-\tau m}\III&=&
  \e^{-\tau (m+1)}(z-X)^{-1}|U_\tau h\rangle\langle U_\tau
    h|(z-X)^{-1}\\
    &\to& (z-X)^{-1}|h_m\rangle\langle h_m|(z-X)^{-1},\\
    \e^{-m\tau}\II+\lambda^{-1}&=&
    \e^{-m\tau}\int_0^\infty h(x)^2\Big((\e^\tau
    z-x)^{-1}+x^{-1}\Big)\d x\\
    &=&\e^{-m\tau}(-z)\e^\tau\int_0^\infty\frac{h(x)^2\d
      x}{x(x-z\e^\tau)}\\
    &=&\e^{-m\tau}\int_0^\infty\frac{h(\e^\tau (-z)y)^2\d
      y}{y(y+1)}\\
    &\to&(-z)^m\int_0^\infty\frac{y^{m-1}\d
      y}{(y+1)}\\
    &=&-(-z)^m\frac{\pi}{\sin\pi(m-1)}\\
    &=&(-z)^m\frac{\pi}{\sin\pi m}.
  \end{eqnarray*}
\qed

\subsection{Point spectrum}

The following theorem is analogous to the characterization of the point spectrum of the Bessel operator described in Theorem 5.2 of \cite{DR}, and is the consequence of the same computation.
    
\bet \ben\item
$w\in\cc\backslash[0,\infty[$ belongs to the point spectrum of
    $H_{m,\lambda}$ iff it satisfies the equation
    \beq
    (-w)^{-m}=\varsigma.\eeq
The corresponding eigenprojection has the kernel
  \beq  \one_{\{w\}}(H_{m,\lambda})(x,y)=\frac{\sin\pi m}{\pi
    m}(-w)^{-m+1}(w-x)^{-1}
  x^{\frac{m}{2}}y^{\frac{m}{2}}(w-y)^{-1}.\eeq
\item $H_0^\rho$ possesses an eigenvalue iff $-\pi<\Im\rho<\pi$, and
  then it is $w=-\e^\rho$. The corresponding eigenprojection has the kernel
    \beq  \one_{\{w\}}(H_{0}^\rho)(x,y)=-w(w-x)^{-1}(w-y)^{-1}.\eeq
  \een  \eet

Let us stress that $\sigma_\p(H_{m,\lambda})$ depends in a complicated way on
the parameters $m$ and $\lambda$.
 There exists a complicated pattern of {\em phase transitions}, when
some eigenvalues ``disappear''. This happens if
\begin{equation}
  \pi\in\Re\frac1m\Ln(\varsigma),\ \
\hbox{ or }\ \ \ 
-\pi\in\Re\frac1m\Ln(\varsigma),\label{excep}
\end{equation}
where $\Ln$ denotes the multivalued logarithm function.
A pair $(m,\lambda)$ satisfying (\ref{excep}) will be called {\em
  exceptional}.
For $m=0$, we need a different condition. We say that $(0,\rho)$ is exceptional if
\beq\Im\rho=-\pi\qquad\text{ or }\qquad\Im\rho=\pi.\eeq

For a given $m,\lambda$ all eigenvalues form a geometric sequence that
lie
on   a logarithmic spiral. This spiral
should be viewed as a curve on the Riemann surface of the logarithm,
and only its ``physical sheet'' gives rise to eigenvalues. For $m$
which are not purely imaginary, only a finite piece of the spiral is
on the ``physical sheet'' and therefore
the number of eigenvalues is finite.

If $m$ is purely imaginary, this spiral degenerates to a
 half-line starting at the origin. Either the whole half-line is on
 the ``physical sheet'', and then the number of eigenvalues is
 infinite, or
the half-line is ``hidden on the
 non-physical sheet of the complex plane'', and then there are no eigenvalues.

 If $m$ is real, the spiral degenerates to a circle. But then the
 operator has at most
 one eigenvalue.

Below we provide a characterization
of $\#\sigma_\p(H_{m,\lambda})$, {\it i.e.}~ of the  number of eigenvalues of $H_{m,\lambda}$. It is proven in \cite{DR}, Proposition 5.3.

\begin{proposition}
Let $m= m_\r+\i m_\i \in \cc^\times$ with $|m_\r|<1$.
\begin{enumerate}
\item[(i)]  Let $m_\r=0$.
  \begin{enumerate}\item[(a)] If $\frac{\ln(|\varsigma|)}{m_\i}\in ]-\pi,\pi[$,
then $\#\sigma_\p(H_{m,\lambda}) = \infty$, \item[(a)] if
 $\frac{\ln(|\varsigma|)}{m_\i}\not \in ]-\pi,\pi[$
then $\#\sigma_\p(H_{m,\lambda}) = 0$.\end{enumerate}
\item[(ii)]  If $m_\r\neq 0$ and if $N\in \nn$ satisfies
$N<\frac{m_\r^2+m_\i^2}{|m_\r|} \leq N+1$, then
\begin{equation*}
\#\sigma_\p(H_{m,\lambda})\in \{N,N+1\}.
\end{equation*}
\end{enumerate}
\end{proposition}

\subsection{M{\o}ller operators}
    
First consider $m\neq0$. We define the M{\o}ller operator
\beq W_{m,\lambda}^\pm:=W^\pm(H_{m,\lambda},X),\label{po1}\eeq
as the operator with  the kernel (\ref{easi}).
Note that
\beq W_{m,\lambda}^{\mp\#}=W^\pm(X,H_{m,\lambda}).\label{po2}\eeq

For $m\neq0$ we have two distinct $\lambda$ with $H_{m,\lambda}$ homogeneous of degree one. One of them is obviously $H_{m,0}=X$. The other is $H_{m,\infty}$. Therefore, the M{\o}ller operators $W_{m,\infty}^\pm$ are functions of $A$.
The M{\o}ller operators $W_{m,\lambda}$ for all $\lambda$ can be expressed in terms of
$W_{m,\infty}^\pm$ and $X$. All this is described in the following theorem:

\bet $W_{m,\infty}^\pm$ exist  {\red as bounded operators} and
\begin{eqnarray}
  W_{m,\infty}^\pm&=&
  \frac{\e^{\mp 2\pi A}+\e^{\mp\i m\pi}\one}
       {\e^{\mp 2\pi A}+\e^{\pm\i m\pi}\one}.
       \end{eqnarray}
Besides, if $(m,\lambda)$ is not exceptional, then
 $W_{m,\lambda}^\pm$ exist {\red as bounded operators}  and are given by
\begin{eqnarray}
    W_{m,\lambda}^\pm&=&\frac{1}{\Big(\one-\varsigma\e^{\pm\i\pi
        m}X^{m}\Big)}\\
    &&-
      W_{m,\infty}^\pm
     \frac{\varsigma\e^{\pm\i\pi
        m}X^{m}}{\Big(\one-\varsigma\e^{\pm\i\pi
        m}X^{m}\Big)}.\label{papa}
\end{eqnarray}
They  satisfy
\begin{eqnarray}
W_{m,\lambda}^{\mp \t} \;\!W_{m,\lambda}^\pm& =& \one,\label{pa1}\\
W_{m,\lambda}^\pm \;\!W_{m,\lambda}^{\mp
  \t}&=&\one_{[0,\infty[}(H_{m,\lambda}),\label{pa2}\\
W_{m,\lambda}^\pm X&=&H_{m,\lambda}W_{m,\lambda}^\pm .\label{pa3}
\end{eqnarray}
\label{thm}\eet

{\red \proof}
By (\ref{easi}), the kernel of $W_{m,\infty}^\pm$ is given by
    \begin{eqnarray}
          W_{m,\infty}^\pm(x,y)&=&\delta(x-y)+\frac{\e^{\pm\i\pi m}\frac{\sin\pi
        m}{\pi}
            x^{-\frac{m}{2}}      y^{\frac{m}{2}}}{x-y\mp\i0}\\
&=&          \delta(x-y)+\frac{\e^{\pm\i\pi m}\frac{\sin\pi
        m}{\pi}}{\sqrt{xy}}
      \Big(\frac{y}{x}\Big)^{\frac{m}{2}}
      \Big(\Big(\frac{x}{y}\Big)^{\frac12}-\Big(\frac{y}{x}\Big)^{\frac12}\mp\i0\Big)^{-1}.\nonumber \end{eqnarray}
    Now, by Subsection \ref{Dilations} and the formula (\ref{integral}), the operator with the kernel
\begin{eqnarray}\frac1{\sqrt{xy}}
      \Big(\frac{y}{x}\Big)^{\frac{m}{2}}      \Big(\Big(\frac{x}{y}\Big)^{\frac12}-\Big(\frac{y}{x}\Big)^{\frac12}\mp\i0\Big)^{-1}\nonumber \end{eqnarray}
   equals
    \beq
    \frac{\pm 2\pi\i}{\e^{\mp\pi\i m\pm2\pi A}+\one}.\eeq
Therefore,
    \begin{eqnarray}
      W_{m,\infty}^\pm&=&\one\pm\e^{\pm\i\pi m}\frac{\sin\pi
        m}{\pi}
      \frac{\pm 2\pi\i}{\e^{\mp\pi\i m\pm2\pi A}+\one}\\
      &=&
  \frac{\e^{\mp 2\pi A}+\e^{\mp\i m\pi}}
       {\e^{\mp 2\pi A}+\e^{\pm\i m\pi}}.
       \label{papa1}
    \end{eqnarray}
To see that (\ref{papa1}) is bounded we use $-1<\Re m<1$.

{\red If $(m,\lambda)$ is not exceptional, then $\frac{1}{(\one-\varsigma\e^{\pm\i\pi
        m}X^{m})}$ and $\frac{\varsigma\e^{\pm\i\pi
        m}X^{m}}{(\one-\varsigma\e^{\pm\i\pi
        m}X^{m})}$ are bounded,
  and therefore the formula (\ref{formu}) defines $W_{m,\lambda}^\pm$ as a bounded operator.
}

(\ref{p1}), (\ref{p2}) and (\ref{p3}) rewritten using (\ref{po1}) and (\ref{po2}) yield
(\ref{pa1}), (\ref{pa2}) and (\ref{pa3}).
\qed

Next, consider $m=0$.  We set
\[W_0^{\rho,\pm}:=W^\pm(H_0^\rho,X).\]
Note that
\[W_0^{\rho,\mp\,\#}=W^\pm(X,H_0^\rho).\]

    For $m=0$, only $X=H_0^\infty$ is homogeneous of degree $1$, therefore we do not have an analog of (\ref{papa}).

    \bet Suppose that  $\rho$ is not exceptional. Then
    the M{\o}ller operators $W_0^{\rho,\pm}$ exist as bounded operators
    and are given by
        \begin{eqnarray}
     W_{0}^{\rho, \pm}&=&\one
\mp     \frac{
       2\pi\i}{(\e^{\pm2\pi A}+\one)}
     \frac{1}{(\ln X\mp\i\pi-\rho)}.
  \label{compact1}      \end{eqnarray}
        They   satisfy
\begin{eqnarray}
W_0^{\rho,\mp \t} \;\!W_0^{\rho,\pm}& =& \one,\\
W_0^{\rho,\pm} \;\!W_0^{\rho,\mp
  \t}&=&\one_{[0,\infty[}(H_0^\rho),\\
W_0^{\rho,\pm} X&=&H_0^\rho W_0^{\rho,\pm} .
\end{eqnarray}
\eet

\proof Using the assumption that
$\rho$ is non-exceptional, {\red we check that $  \frac{1}{(\ln X\mp\i\pi-\rho)}$ is bounded.
  Now, the formula (\ref{compact}) proves (\ref{compact1})
  and the boundedness of
$     W_{0}^{\rho, \pm}$. } \qed

    \section{Equivalence with Schr\"odinger operators with inverse square potentials}

    \setcounter{equation}{0}
\label{Equivalence}

        Recall that in \cite{BDG}
        a holomorphic family of  closed operators
        $H_m$ was introduced.
        We change slightly the notation for these operators, and we will denote them by $\tilde H_m$ in this paper.

        Thus, for $\Re m>-1$,  $\tilde H_m$
        is the unique closed operators on $L^2[0,\infty[$ given on $C_{\rm c}^\infty]0,\infty[$ by the differential expression
    \beq \tilde H_m=-\partial_x^2+\Big(-\frac14+m^2\Big)\frac{1}{x^2},\label{schro1}\eeq
    such that functions in its domain behave as $x^{\frac12+m}$ around zero.
    $\tilde H_m$ can be  diagonalized
    with help of the so-called Hankel transformation $\cF_m$, which is a bounded invertible involutive operator such that
        \begin{eqnarray}\cF_m\tilde H_m\cF_m^{-1}&=&X^2,\label{f1}\\
          \cF_mA\cF_m^{-1}&=&-A.\label{f2}
          \end{eqnarray}

        For $-1<\Re m<1$, two more general family of operators
     $H_{m,\kappa}$  and $H_0^\nu$ were constructed in \cite{DR}.
        In this paper they will be be denoted by $\tilde H_{m,\kappa}$
        and $\tilde H_0^\nu$.

        $\tilde H_{m,\kappa}$ is given by  the differential expression on the right hand side of (\ref{schro1}) with the boundary condition at zero $\kappa x^{\frac{1}{2}-m}+ x^{\frac{1}{2}+m}$.
 $\tilde H_0^\nu$ is defined by (\ref{schro1}) with $m=0$ and the boundary conditions $x^{\frac{1}{2}}\log(x)+\nu x^{\frac{1}{2}}$.
Note that
\[ \tilde H_m=\tilde H_{m,0}=\tilde H_{-m,\infty},\quad \tilde H_0=\tilde H_0^\infty.\]

    Define the unitary operator
    \beq (If)(x):=x^{-\frac14}f(2\sqrt x).\eeq
    Its inverse is
        \beq
        (I^{-1}f)(x):=\Big(\frac{y}{2}\Big)^{\frac12}f\Big(\frac{y^2}{4}\Big).\eeq
        Note that
        \begin{eqnarray}
          I^{-1}XI&=&\frac{X^2}{4},\label{x1}\\
          I^{-1}AI&=&\frac{A}{2}.\label{x2}
        \end{eqnarray}

        \bet
        We have
        \beq
        \cF_m^{-1}I^{-1} H_{m,\lambda}I\cF_m=\frac14\tilde
        H_{m,\kappa},\label{qeq}\eeq
        where the pairs $(m,\lambda)$ and $(m,\kappa)$ are linked by
        the relation
        \beq
\lambda\frac{\pi}{\sin(\pi m)}=\kappa\frac{\Gamma(m)}{\Gamma(-m)  },\label{varsigma}\eeq
(The relation (\ref{varsigma}) is equivalent to saying that
the parameter $\varsigma$ introduced in \cite{DR} (5.2) coincides with the
parameter
$\varsigma$ introduced in (\ref{var}).)\label{link1}
\eet

\proof To avoid notational collision, we denote by $\tilde W_{m'm}^\pm$
the M{\o}ller operators, denoted by $W_{m,m'}^\pm$ in \cite{DR}.
We quote some identities from \cite{DR} contained in  Prop. 4.11, Prop. 4.9 and Equation (6.3):
\begin{eqnarray}
  \tilde  W_{-m,m}^\pm&=&\frac{\e^{\pm\pi A}+\e^{\mp\i\pi m}}{\e^{\pm\pi
      A}+\e^{\pm\i\pi m}},\label{a1}\\
    \big(k^2+\tilde H_{-m}\big)^{-1}&=&
    \tilde  W_{-m,m}^\pm \big(k^2+\tilde H_{m}\big)^{-1}\tilde  W_{-m,m}^{\pm-1},
    \label{a2}
\\ 
\big(k^2+\tilde H_{m,\kappa}\big)^{-1}&=&\frac{1}{1-\varsigma(\frac{k}{2})^{2m}}
\big(k^2+\tilde H_{m}\big)^{-1}\nonumber\\&&
-  
\frac{\varsigma\big(\frac{k}{2}\big)^{2m}}{1-\varsigma(\frac{k}{2})^{2m}}
\big(k^2+\tilde H_{-m}\big)^{-1}.\label{a3}
\end{eqnarray}

On the other hand, by Theorem \ref{thm} we have
\begin{eqnarray}
W_{m,\infty}^\pm&=&
  \frac{\e^{\mp 2\pi A}+\e^{\mp\i m\pi}\one}
       {\e^{\mp 2\pi A}+\e^{\pm\i m\pi}\one},\label{b1}  \\
  (z-H_{m,\infty})^{-1}&=&W_{m,\infty}^\pm(z-X)^{-1}W_{m,\infty}^{\pm-1},\label{b2}
\\
  (z-H_{m,\lambda})^{-1}&=&\frac{1}{1-\varsigma(-z)^m}(z-X)^{-1}  \nonumber\\&&
  -\frac{\varsigma(-z)^m}{1-\varsigma(-z)^m}(z-H_{m,\infty})^{-1}.
\label{b3}
\end{eqnarray}

Now (\ref{f2}), (\ref{x2}), (\ref{a1}) and (\ref{b1}) imply
\begin{eqnarray}
  \cF_m^{-1}I^{-1}W_{m,\infty}^\pm I\cF_m&=&\tilde  W_{-m,m}^\pm.\label{c1}
\end{eqnarray}
Setting $ -z=\frac{k^2}{4},$ using (\ref{b2}), (\ref{c1}), (\ref{f1}), (\ref{x1}) and (\ref{a2})
we check that
\begin{eqnarray}
    \cF_m^{-1}I^{-1}(z-H_{m,\infty})^{-1} I\cF_m&=&\Big(z-\frac{\tilde
      H_{-m}}{4}\Big)^{-1}.
\end{eqnarray}
Finally, (\ref{a3}) and (\ref{b3}) yield
\begin{eqnarray}
        \cF_m^{-1}I^{-1}(z-H_{m,\lambda})^{-1} I\cF_m&=&\Big(z-\frac{\tilde
      H_{m,\kappa}}{4}\Big)^{-1}.
\end{eqnarray}
\qed

\bet We have
\beq
        \cF_0^{-1}I^{-1} H_0^\rho I\cF_0=\frac14\tilde
        H_0^\nu,\label{qeq2}\eeq
where $\rho=-2\nu$. \label{link2}\eet

\proof
The $m=0$ case will be reduced to the $m\neq0$.

First note that the family
\beq (m,\nu)\mapsto\begin{cases}\tilde H_{m,\frac{m\nu-1}{m\nu+1}},& m\neq0\\
\tilde H_0^\nu,
&m=0;\end{cases}\eeq
is analytic
(see \cite{DR}, Remark 2.5).
Hence,  for $m\to0$,
\beq \tilde H_{m,-1+2m\nu }\to \tilde H_0^\nu.\eeq

Similarly,
by Theorem \ref{annal1}, the family
\beq (m,\rho)\mapsto\begin{cases} H_{m,\frac{m}{1-m\rho}},& m\neq0\\
H_0^\rho,&m=0\end{cases}\eeq
is analytic. Hence, for $m\to0$,
\beq H_{m,m+m^2\rho}\to H_0^\rho.\eeq

Around $m=0$, the condition
(\ref{varsigma}) becomes
\beq \frac{\lambda}{m}\simeq-\kappa\big(1+O(m)\big).\eeq
Therefore, around $m=0$, (\ref{qeq}) implies
        \beq
        \cF_m^{-1}I^{-1} H_{m,m+m^2\rho}I\cF_m=\frac14\tilde
        H_{m,-1+2m\nu+O(m^2)},\label{qeq1}\eeq
where  $\rho=-2\nu$. Passing to the limit   $m\to0$ in (\ref{qeq1}) we obtain
(\ref{qeq2}). \qed
    \appendix
    \section{Some integrals}
    \setcounter{equation}{0}

    {\red \begin{itemize}
   \item
    For $-\frac{1}{2}<\Re a<-\frac{1}{2}$,
    \beq
    \int_{-\infty}^\infty\frac{\e^{at}\d
      t}{\e^{\frac{t}{2}}+\e^{-\frac{t}{2}}}
    =\frac{\pi}{\cos\pi a}.\label{integral0}
    \eeq
    Indeed, we integrate the analytic function
    $f(t):=\frac{\e^{at}
    }{\e^{\frac{t}{2}}+\e^{-\frac{t}{2}}}$ over the rectangle
    $-R,R,R+2\pi\i,-R+2\pi\i$,  use
    $f(t\pm 2\pi \i)=-\e^{2\pi\i a}f(t)$ and $\Res f(\pi\i)=\frac{1}{\i}\e^{\pi\i a}$.
   \item
    For the same $a$,
    \beq
    \int_{-\infty}^\infty\frac{\e^{at}\d
      t}{\e^{\frac{t}{2}}-\e^{-\frac{t}{2}}\mp\i0}
    =\frac{\pm 2\pi\i}{\e^{\pm 2\pi\i a}+1}.\label{integral}
    \eeq
To see this we shift the integration in (\ref{integral0}) from $\rr$ to $\rr\pm\i\pi$ without crossing the singularity at $\pm\pi\i$.
  \item
    Setting $a=-\i\xi$ in (\ref{integral}) we obtain the following Fourier transform:
    \beq h(t)=\frac{1}{\big(\e^{\frac{t}{2}}-\e^{-\frac{t}{2}}\mp\i0\big)},\quad
    \hat h(\xi)=\frac{\pm 2\pi\i}{(\e^{\pm2\pi\xi}+1)}.\label{integral1}\eeq
 \item
      For $-1<\Re m<0$,
      \beq\int_0^\infty s^{m}(1+s)^{-1}\d s=-\frac{\pi}{\sin \pi m}.\label{app4}
      \eeq
Indeed, we set $m=a-\frac12$ and $s=\e^t$ in (\ref{integral0}).
    \end{itemize}
}

\end{document}